\documentclass{mem}
\usepackage{natbib}\usepackage{txfonts}\usepackage{balance}
\usepackage{graphicx}
\usepackage{epstopdf}
\usepackage[a4paper,breaklinks,dvipdfm]{hyperref}
\idline{75}{282}
\begin{document}
\def\teff{$T\rm_{eff }$}
\def\kms{$\mathrm {km s}^{-1}$}
\def\mdot{$m\rm_{dot }$ }
\def\msun{M$_{\odot} $ }
\title{The X-ray emission of magnetic cataclysmic variables in the XMM-Newton era}

   \subtitle{}

\author{
M. \,Mouchet\inst{1,2}, 
 J.M. \,Bonnet-Bidaud\inst{3}
\and D. \,de Martino\inst{4}          }

  \offprints{M. Mouchet}

\institute{
Laboratoire Astroparticule et Cosmologie,
Universit\'e Paris Diderot --
F-75013 Paris, France \, \email{martine.mouchet@apc.univ-paris-diderot.fr}
\and
Laboratoire LUTH, 
Observatoire de Paris, CNRS, Universit\'e Paris-Diderot, F-92190 Meudon, France \and
CEA Saclay, DSM/Irfu/Service d'Astrophysique, F-91191 Gif-sur-Yvette, France
\and
INAF Capodimonte Observatory, Salita Moiariello 16, I-80131 Napoli, Italy
}

\authorrunning{Mouchet}

\titlerunning{X-ray emission of MCVs}

\abstract{
We review the X-ray spectral properties of magnetic cataclysmic binaries 
derived from observations obtained during the last decade with the large X-ray 
observatories XMM-Newton, Chandra and Suzaku. We focus on the signatures of the
different accretion modes which are predicted according to the values of the
main physical parameters (magnetic field, local accretion rate and 
white dwarf mass).
The observed large diversity of spectral behaviors indicates a wide range of
parameter values in both intermediate polars and polars, in line with a
possible evolutionary link between both classes.
  
}
\maketitle{}

\section{Introduction}
In magnetic cataclysmic binaries, the magnetic field of the white dwarf is 
strong enough to channel the accreting matter along the field lines from the 
magnetosphere. 
Material thus falls down close to the polar caps, along an accretion column 
or curtain. 
Two main categories are distinguished: the synchronous Polars (AM-Her type) 
and the non-synchronized Intermediate Polars (IPs) (P$_{spin}<$P$_{orb}$).\\
More than one hundred polars are presently known (Ritter \& Kolb (2003) 
catalog, version June 2011), having orbital/spin periods ranging from 77\,min. 
to 14\,h with more than a half having a period below the evolutionary gap 
(below 2h). 
Their magnetic fields  have been determined by different methods (cyclotron
lines, Zeeman components, optical polarimetry) and their values range  between 
7 and 230\,MG (Wickramasinghe \& Ferrario 2000, G\"ansicke et al. 2004).
Polars often exhibit alternate high and low states, associated with variations 
of the accretion rates (10$^{-10}$ to 10$^{-13}$ \msun yr$^{-1}$). 
For seven of them, the white dwarf rotation differs slightly  from the orbital 
one by a few percent. \\
IPs are less numerous (36 secure and 20 probable in Mukai (2011) catalog, 
version Jan. 2011). 
Their orbital periods range from 80\,min. to 48\,h, with only 5 objects below 
the gap, and with spin-to-orbital period ratios ranging from 9x$10^{-4}$
to 0.68. 
For eight sources, polarized ~ optical flux has been detected with 
estimates of the magnetic field  between 8 to 15 MG,  except for V405\,Aur  
(30 MG) (Katajainen et al. 2010).
The accretion occurs directly from the disc (disc-fed), possibly with  a 
stream overflow and, for at least one source (V2400 Oph), with no disc 
(Hellier 2007). 
The accretion geometry strongly depends on the combinaison of the accretion 
rate and B values (Norton et al. 2008).\\
As already extensively debated, whether or not IPs are progenitors of polars,
with similar magnetic field values, is still an open question.

\section{The accretion models}
\subsection{The standard shock model}
The basic picture of the accretion onto a magnetic white dwarf consists in  
a field aligned flow directed to the magnetic poles which is shocked before 
reaching the WD surface, with most of the gravitational energy  released 
in the post-shock region (PSR).  
The resulting shock temperature is given by $T_s=3GMm_h/8kR$, of the order 
of 10 to 50\,keV (M and R being the WD mass and radius). 
The corresponding X-ray luminosity $L_X = G M M_{dot}/R$ is of the order 
of $10^{31}-10^{33}$ erg s$^{-1}$ ($M_{dot}$ being the accretion rate).
Below the shock, bremsstrahlung and cyclotron radiation are the two main 
competitive cooling processes.  Hard X-rays (E$>$2\,keV) are radiated 
from the post-shock plasma as well as optical/IR cyclotron. 
Half of the accreting luminosity  illuminates the polar caps and is 
reprocessed as soft X-ray blackbody emission (E$<$2\,keV) with a temperature 
$T_{bb}=(L_{acc}/f\,8\pi R^2 \sigma)^{1/4}$, of the order of 10 to 50\,eV (f being the irradiated fractional area of the stellar surface).

An exhaustive review of the hydrodynamic steady  and non-stationary models of 
the post-shock region emission has been presented by Wu (2000).
Since then, further developments have been proposed. 
They include, for instance, the 2-temperature  models, required for strong 
magnetic fields, low accretion rates and high WD masses (Fischer \& Beuermann 
2001, Saxton et al. 2005, Imamura et al. 2008), 
the effect of mass leakage and heating at the WD surface (Wu et al. 2001),
the inclusion of the dipole field geometry (Canalle et al. 2005, 
Saxton et al. 2007), or  the generalization of the analytical solutions 
(Lamming 2004, Falize et al. 2009) (see also C. Michaut in these proceedings).
These studies provide the temperature and density profiles of the post-shock 
region  in the accretion column which in turn allow  us to predict the 
X-ray spectra, via the use of optically thin plasma 
codes.
 
\subsection{Three main accretion regimes}
\begin{figure}[]
\resizebox{\hsize}{!}{\includegraphics[clip=true]{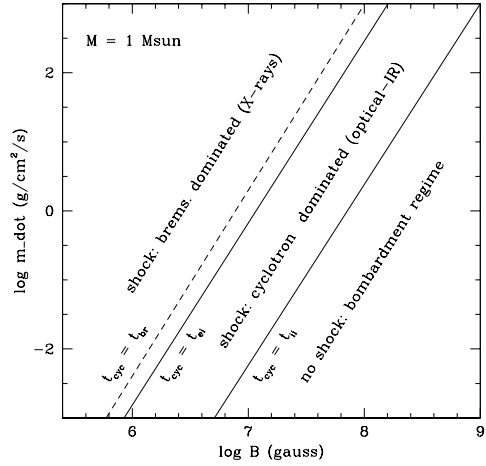}}
\caption{\footnotesize
Regions in the specific accretion rate $m_{dot}$- B plane of the three 
accretion regimes (adapted from Lamb \& Master 1979).
}
\label{mdot-B}
\end{figure}

As originally discussed by King \& Lasota (1979) and Lamb \& Master (1979), the presence of 
a stand-off shock strongly depends on the specific local accretion rate \mdot 
and the  magnetic field value B.  
Three main regimes can be distinguished (see Fig.1): 
a) a {\it standard shock} is formed above the WD surface for high \mdot and 
low B, 
b) a  {\it blobby accretion} is expected for very high \mdot 
(Kuijpers et al. 1982,  Frank et al. 1988) for which shocks are buried 
below the 
WD photosphere, leading to a strong soft X-ray blackbody component  
and c) a  {\it bombardment regime} is associated with very low \mdot and 
high B (Kuijpers et al. 1982 ,Woelk \& Beuermann 1996 and ref. herein, Fischer
and Beuermann 2001). In this last case,
Coulomb ion-ion interaction time been longer than the cooling time, 
this prevents the formation of a shock. 
A low temperature X-ray emission is expected for this regime.  

X-ray continuum and lines provide powerful diagnostics to probe the 
temperature and density structures of the
accretion column which can be confronted to the predictions of the different
regimes.
Below we present examples of these three regimes which are encountered in 
magnetic CVs. 
Their properties are derived from detailed X-ray observations of IPs and 
polars, which are now accessible thanks to the 
new generation of X-ray satellites (XMM-Newton, Chandra and Suzaku) 
available during the last decade.

\section{The accretion modes in Intermediate Polars}
Considering first the intermediate polars,
 among 36 secure IPs identified at present time, 30 were observed with 
XMM-Newton, 10 with Chandra and 22 with Suzaku. 
Eleven of them have been confirmed as IPs using XMM-Newton data which 
revealed their spin period (see de Martino et al. 2008, Anzolin et al. 2008, 
Bonnet-Bidaud et al. 2009, Anzolin et al. 2009, de Martino et al. 2009, 
for the most recent identifications). 
Numerous IPs are hard X-ray (E$>$20\,keV) sources; 18 were detected with 
INTEGRAL (Scaringi et al. 2010) and 22 with Swift/BAT (Brunschweiger et al. 
2009). 
Notably the Galactic ridge spectrum is
similar to the hard X-ray spectra of IPs (Revnivtsev et al. 2009), indicating 
that they are strong potential contributors to the diffuse galactic 
hard X-ray emission. 
 The study of their X-ray spectra has thus benefited 
from their detection at higher energies.

\subsection{Hard-X-ray emission and stand-off shocks}
For most IPs the cooling below the shock is expected to be dominated by 
bremsstrahlung radiation. The exact solution of a stationary post-shock region 
in 1D plane-parallel geometry has been derived by Revnivtsev et al.(2004) and 
found to be very similar to the analytical solution for a  constant pressure 
medium.
 The temperature and density profiles depend only slightly from the specific 
accretion rate and thus, the resulting X-ray spectral shape leads to a direct 
evaluation of the WD mass (Suleimanov et al. 2005, Yuasa et al. 2010). 
A possible contribution of Compton up-scattering in the post-shock region 
has also been evaluated by Suleimanov et al.(2008) but its effect is 
only significant
for high WD masses. 
The most recent mass determinations using Suzaku data of 17 IPs 
(Yuasa et al. 2010, see also these proceedings) have been obtained by coupling
the PSR structure computed by Suleimanov et al. (2005) with the APEC plasma 
code (Smith et al. 2001). 
The derived average mass is 0.88 $\pm$ 0.25 \msun, slightly higher than 
previous estimates (Suleimanov et al. 2005, Brunschweiger et al. 2009). 

\subsection{Soft X-ray IPs}
As mentioned above, the standard shock model predicts a reprocessed component 
arising from the illumination of the WD surface by the hard X-ray component. 
Contrary to polars, the first discovered IPs did not show such a component; 
this was 
interpreted as due to either a strong absorption in the accretion curtain or 
to a lower  temperature EUV emission associated to a larger irradiated zone.
The number of IPs detected in soft X-rays (E$<$1\,keV) has increased from 
4 objects seen with ROSAT up to 13 observed with XMM-Newton (about 40\%) 
(Evans \& Hellier 2007, Anzolin et al. 2008). 
 When the soft-X-ray component is fitted with a blackbody spectrum (see an exemple in Fig.2),
the resulting  temperatures (typically 50-120\,eV) are higher than typically 
found in polars (20-60\,eV) with a lower soft/hard flux ratio (Fig. 3).
Yet for polars the accreting surfaces are expected to be smaller than 
for IPs, implying a higher temperature for a given accretion rate.  The higher 
temperatures found in IPs would indicate higher accretion rates, 
in line with higher luminosities. 
The hard X-ray IP emitters being expected to have an undetected reprocessed 
component shifted in the EUV, the absence of IPs with temperatures similar 
to those of polars is striking. 
Such a dichotomy can be related to the magnetic field. 
Indeed the fact that seven over eight IPs with detectable circular 
polarisation require a soft X-ray component to be fitted, as  for polars,
is in favor of the evolutionary link from IPs to polars of similar B values, 
at least for this sub-class of soft IPs. 
Searches for polarized flux in all soft IPs are strongly needed.    
\begin{figure}[]
\resizebox{\hsize}{!}{\includegraphics[clip=true]{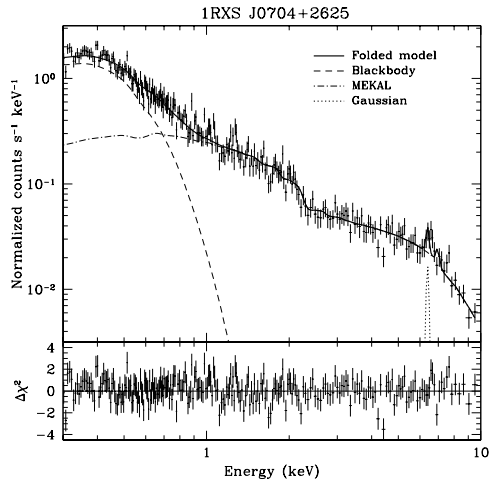}}
\caption{\footnotesize
XMM-Newton EPIC spectrum of the IP RXS J070407.9+262501 fitted with a blackbody soft
component (dotted line, T$_s$=84\,eV) and an optically thin plasma component (dashed-dotted line, T$_h$=44\,keV). A gaussian component is added to fit the fluorescence Fe  K$_{\alpha}$ line (from Anzolin et al. 2008).
}
\label{rx1730spe}
\end{figure}
\begin{figure}[]
\resizebox{\hsize}{!}{\includegraphics[clip=true]{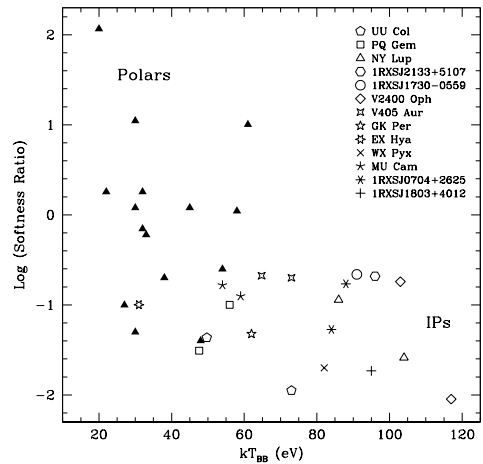}}
\caption{\footnotesize
Soft X-ray -to-hard X-ray ratios of magnetic CVs (from Anzolin et al. 2008).
Values for polars are taken from Ramsay et al. 2004.
}
\label{ratio}
\end{figure}

\subsection{Additional features of the X-ray spectra}
X-ray spectra of IPs are generally well described in the context of the 
stand-off shock model predicting both hard X-ray and soft X-ray reprocessed 
emission.
However some sources show more complex spectra, exhibiting signatures of 
partial cold and/or warm absorbers, as it is expected when the line of sight 
intercepts the accretion curtain.
Also the presence of a  iron K$_{\alpha}$ line (EW$\sim$ 100\,eV) associated
 with a reflection component onto the surface of the white dwarf or in 
the accretion flow  is  often detected.
 A typical exemple of such a complex spectrum is found in the IP 
1RXS J173021.5-055933 (de Martino et al. 2008). 
The inclusion of these additional contributions affects the estimate of the 
WD masses (Hayashi et al. 2011).

\section{The accretion modes in Polars}
Turning now to polars,
among the 104  reported in Ritter \& Kolb (2003) catalog,  
60 have been observed so far with XMM-Newton, 9 with Chandra and 2 
with Suzaku.
Ramsay \& Cropper (2004) and Ramsay et al. (2004) performed  XMM-Newton 
snapshot 
series of 37 polars; surprisingly 16 were found to be in a low state and 
among the 21 in a high state, 6 do not require a soft X-ray component to 
fit their spectra. 
This fraction was confirmed with further observations analysed by 
Ramsay et al. (2009): among 27 polars in a high state, 10 do not exhibit 
any soft component.

Contrary to IPs, only 4 polars are detected at high energies with INTEGRAL, 
including the two 
 asynchronous systems BY Cam and  V1432 Aql (Landi et al. 2009, 
Scaringi et al. 2010). 
This is in agreement with the lower accretion rates in polars compared to IPs.

\subsection{Low states and the bombardment regime}
During optical low states, polars are weak X-ray emitters.
Thanks to the great sensitivity of XMM-Newton a precise description of the 
corresponding spectra is now accessible. 
They are two possible X-ray emission contributors : either flux from a 
residual weak accretion  or a coronal emission from the companion star.  
One of the best documented polars in low state is EF Eri which stays in this 
state since ten years (Schwope et al. 2007). 
While it is a hard X-ray source during high states, its low state spectrum
is well described wih an optically thin plasma at a temperature of 2.8 keV. 
The soft X-ray luminosity (2x10$^{29}$ erg s$^{-1}$) is compatible with a 
coronal emission of the cool companion, but the detection of IR-optical 
cyclotron emission indicates that accretion is still occuring. 
The ratio of the cyclotron and X-ray emission of the order of 60 is in rough 
agrement with the bombardment solution computed by Beuermann (2004) for a 
magnetic field value of 14\,MG and a weak specific accretion rate of 10$^{-2} 
$g\,cm$^{-2}$\,s$^{-1}$.
In addition X-ray flares have been detected during low states of some polars, 
among them VV Pup, V393 Pav (Pandel \& Cordova 2002) or UZ For (Still \& Mukai
2001). 
Coronal ejections close to the inner Lagrangian point of a 10$^{17}$g mass 
can account for the flare intensities.

The bombardment regime has also been claimed to occur in pre-polars. 
These sources are characterized  by a high magnetic field value (derived from 
the position of strong cyclotron features), 
a very low accretion rate (\mdot$<10^{-2}$\,g\,cm$^{-2}$\,s$^{-1}$,  
a cool white dwarf (T$<$10000K) and an orbital period larger than 3h,
except for one among 10 known objects (Schwope et al. 2009 and 
references herein). 
Only 4 are detected in X-rays as weak sources:
SDSS J155331.12+551614.5 and SDSS J132411.57+032050.5 (Szkody et al. 2004), 
WX LMi  (Vogel et al. 2007) and HS0922+1333 (Vogel et al. 2011).
Their X-ray spectra are compatible with one or two optically thin plasmas
(temperature of about 0.3 to 1\,keV),
 associated with  accretion in a bombardment mode from a stellar wind and 
with an eventual coronal emission of 
 a late type star companion in the saturated regime, corresponding to 
a constant X-ray to-bolometric luminosity ratio independent of the rotational 
period (Pizzolato et al. 2003). 
 These detached systems are thought to be in a post-common envelope phase 
prior to a Roche lobe overflow.
         
\subsection{Soft X-ray excess and the blobby accretion}
During high states, a majority of polars exhibit a soft X-ray component which 
luminosity can be much higher than the hard X-ray component 
(ratio L$_{soft}$/L$_{hard}>3$ for 5 among 21 sources (Ramsay \& Cropper 2004).
 An extreme case is V1309 Ori with a ratio greater than 6700 
(Schwarz et al. 2005). 
Two other exemples recently described are AI Tri (Traulsen et al. 2010) and
 QS Tel (Traulsen et al. 2011). 
The corresponding X-ray light curves exhibit strong flickering. 
 These X-ray properties are consistent with an inhomogeneous flow; dense 
blobs of matter sink deeply into the WD atmosphere, shock are buried, 
preventing the detection of a hard X-ray component.   
It is also noteworthy that all sources mentioned above, except one, have a 
magnetic field larger than 40\,MG. 
This is fully consistent with a cyclotron dominated post-shock region. Note 
however that using the larger sample of Rosat observations, Ramsay \& Cropper 
(2004) did not find a trend of energy balance ratio with the magnetic field.

\subsection{Hard X-ray polars and stand-off shock}
In the analysis of XMM-Newton polar observations  made by Ramsay et al. (2009),
 ten sources among the 27 found in a high state show a hard component only
 (for BY Cam and CD Ind, the soft component is absent for one pole only).
Their light curves do not show strong flickering. 
This tends to favor the standard shock model which prevails in a large
number of IPs, with a dominating hard X-ray bremsstrahlung contribution and a  
reprocessed component shifted at lower energies (EUV). 
This could be explained if these ten sources harbor lower magnetic fields 
implying larger accreting surfaces,  but the B values of this sample 
is not yet well documented. 
Note that among the ten polars mentioned above, three are asynchronized 
systems (V1500 Cyg, BY Cam, CD Ind), renforcing the analogy with IPs.    

\section{X-ray line diagnostics}
For the brighest objects, high-resolution X-ray spectroscopy is now 
available using Chandra and XMM-Newton gratings. 
It reveals a lot of emission lines with a large range of ionisation potentials (see review by Ishida 2010), in agreement with the stratification in temperature and density of the PSR.
The usual diagnostics for the temperature of the emitting zone is 
obtained from the He-like to H-like line ratio.
The He-like triplet provides ratios sensitive to the  density, however these ratios are affected by UV photo-excitation. 
To circumvent this difficulty, Mauche et al. (2003) have proposed the 
Fe XXII line ratio 
(11.92\AA/11.77\AA) which has the advantage to have a high critical density 
of the order of 5 10$^{13}$  cm$^{-3}$ and to be  insensitive to temperature and photo-excitation. 
They have applied this diagnostic to Chandra observations of EX Hya and found 
a high PSR plasma density n $> 2\,10^{14}$cm$^{-3}$. 
Using such line diagnostics, Girish et al. (2007) could map the temperature 
and density profiles in the post-shock region of AM Her, implying a  
$\sim$1\,\msun  white dwarf. This PSR structure is in agreement with the line 
velocity  stratification derived from ions of different ionisation stages.

\section{Concluding remarks}

During the last decade, the new generation of X-ray observatories has given access to a better knowledge of the accretion processes in magnetic CVs. 
Although a few objects have been discovered in their size-limited sky surveys,
their high sensitivity has  permitted the confirmation of the magnetic 
nature of a large number of candidates found in either hard X-rays or 
optical surveys.
This enlarged sample reveals an increased number of soft X-ray IPs
 and of hard X-ray polars, indicating, for both classes, a wide range 
of values for the main physical parameters (magnetic field, accretion rate,
spin and mass of the WD). 
The general X-ray spectral characteristics are roughly in agreement with the 
different accretion modes expected for specific ranges of these parameters.

However the confrontation of the observations to the models is often done 
using time-averaged spectra, preventing one from resolving fine viewing-angle 
effects.
High temporal resolution spectra is only available for a very few sources (e.g. EX Hya and FO Aqr,  Pek\"on \& Balman 2011a, b), giving access to detailed studies of the orbitally and spin phase modulated absorption effects. 

In the framework of the stand-off shock model, thermal instabilities 
are predicted together with quasi-periodic oscillations (QPOs) 
 depending on the cooling flow functions (Langer et al., 1981, 
Chevalier \& Imamura 1982). Such QPOs  on a timescale of a 
few seconds are
indeed detected in the optical flux of five polars but  
only upper limits have been reported so far in the X-rays (Imamura et al. 2000 and references herein).
 A new approach of this phenomenon is proposed, consisting in the production of a shock in laboratory and the subsequent search for oscillations 
(see C. Michaut in these proceedings).  
  
On an observational point of view, the future X-ray satellites will have 
similar temporal and spectral resolving capabilities as in the optical, 
allowing , for instance, tomographic studies of the accreting regions of 
the white dwarf. Also, detailed atomic features will be accessible thanks to 
the X-ray calorimeter aboard 
Astro-H with a resolution of 4\,eV at 6\,keV, such as 
the dielectronic satellite lines of He and H-like iron lines which are additional accurate diagnostics of the physical parameters of the corresponding emitting regions (Ishida 2010).

\begin{acknowledgements}
M.M. acknowledges financial support by the GDR PCHE (CNRS) and D.d.M. 
 by ASI INAF Contract N. I/009/10/0.   
\end{acknowledgements}

\bibliographystyle{aa}

\begin{thebibliography}{}

\bibitem[{Anzolin et al.(2008)}]{anzolin08}
Anzolin, G.~et al. 2008, A\&A, 489, 1243

\bibitem[{Anzolin et al.(2009)}]{anzolin09}
Anzolin, G.~et al. 2009, A\&A, 501, 1047
 
\bibitem[{Beuermann (2004)}]{beuermann04}
Beuermann, K. 2004, 
 IAU Colloquium 190, 
Eds S. Vrielmann and M. Cropper. ASPC Proceedings,  315, 187


\bibitem[{Bonnet-Bidaud et al.(2009)}]{bb09} 
Bonnet-Bidaud, J.-M., de Martino, D., Mouchet, M. 2009, ATel 1895 

\bibitem[{Brunschweiger et al.(2009)}]{brunschweiger09}
Brunschweiger, J.~et al. 2009,  A\&A, 496, 121

\bibitem[{Canalle et al.(2005)}]{canalle05}
Canalle, J.B.G.~et al. 2005,  A\&A, 440, 185

\bibitem[{Chevalier \& Imamura(1982)}]{chev82}
Chevalier, R.A.\& Imamura, J.N. 1982, \apj, 261, 543

\bibitem[{Evans \& Hellier(2007)}]{evans07}
Evans, P.A., Hellier, C. 2007, \apj, 663, 1277

\bibitem[{Falize et al.(2009)}]{falize09}
Falize, E.~et al. 2009, Ast.~Space~Sci., 322,~71

\bibitem[{Fischer \& Beuermann(2001)}]{fischer01}
Fischer, A. \& Beuermann, K. 2001,  A\&A, 373, 211

\bibitem[{Frank et al.(1988)}]{frank88}
Frank, J., King, A.R., Lasota, J.P. 1988, A\&A, 193, 113  

\bibitem[{G\"aensicke et al.(2004)}]{gaensicke04}
G\"aensicke, B.~et al. 2004, \apj, 613, L141

\bibitem[{Girish et al.(2007)}]{girish07}
Girish V., Rana, V.R., \& Singh, K.P. 2007, \apj, 658, 525

\bibitem[{Hayashi et al.(2011)}]{hayashi11}
Hayashi, T.~et al. 2011, astro-ph/1106.611v1

\bibitem[{Hellier(2007)}]{hellier07}
Hellier, C. 2007, Proc. IAU Symp. 243, 
eds J. Bouvier \& I. Appenzeller, p325 

\bibitem[{Imamura et al.(2000)}]{ima00}
Imamura, J.N., Steiman-Cameron, T.Y., Wolff, M.T. 2000, PASP, 112, 18

\bibitem[{Imamura et al.(2008)}]{ima08}
Imamura, J.N., Bryson, W.C., Steiman-Cameron, T.Y. 2008, PASP, 120, 171

\bibitem[{Ishida(2010)}]{ishida10}
Ishida, M. 2010, Space Sci. Rev., 157, 155

\bibitem[{Katajainen et al.(2010)}]{kata10}
Katajainen, S.~et al. 2010, \apj, 724, 165

\bibitem[{King \& Lasota(1979)}]{king79}
King, A.R., \& Lasota, J.P. 1979, \mnras, 188, 653

\bibitem[{Kuijpers \& Pringle(1982)}]{kuijpers82}
Kuijpers, J., Pringle, J.E. 1982,  A\&A, 114, L4

\bibitem[{Lamb \& Master(1979)}]{lamb79}
Lamb, D.Q., Master, A.R. 1979, \apj, 234, L117

\bibitem[{Laming(2004)}]{laming04}
Laming, J.M. 2004, Phys. Rev. E, 70, 057402

\bibitem[{Landi et al.(2009)}]{landi09}
Landi, R.~et al. 2009,  A\&A, 392, 630

\bibitem[{Langer et al.(1981)}]{langer81}
Langer, S.H., Chanmugam, G., Shaviv, G. 1981, \apj, 245, L23

\bibitem[{de Martino et al.(2008)}]{ddm08}
de Martino, D.~et al. 2008, A\&A, 481, 149

\bibitem[{de Martino et al.(2009)}]{ddm09}
de Martino, D.~et al. 2009, ATel 2089

\bibitem[{Mauche et al.(2003)}]{mauche03}
Mauche, C.W., Liedahl, D.A., Fournier, K.B. 2003,  \apj, 588, L101 


\bibitem[{Mukai(2011)}]{mukai11}
Mukai, K. 2011, http://asd.gsfc.nasa.gov/\\
Koji.Mukai
/iphome/catalog/alpha.html

\bibitem[{Norton et al.(2008)}]{norton08}
Norton, A.J.~et al. 2008, \apj, 672, 524


\bibitem[{Pandel \& C\'ordova(2002)}]{pandel02}
Pandel, D., C\'ordova, F.A. 2002, \mnras, 336, 1049


\bibitem[{Pandel \& C\'ordova(2005)}]{pandel05}
Pandel,~D.,~C\'ordova,~F.A. 2005,~\apj,~620,~416

\bibitem[{Pek\"on \& Balman(2011a)}]{pekon11a}
Pek\"on, Y. \& Balman, S. 2011a, The X-ray Universe 2011, \\
http://xmm.esac.esa.int/external/xmm\_science/\\
workshops/2011symposium/article id.266

\bibitem[{Pek\"on \& Balman (2011b)}]{pekon11b}
Pek\"on, Y. \& Balman, S. 2011b, \mnras, 411, 1177

\bibitem[{Pizzolato et al.(2003)}]{Pizzo03} 
Pizzolato, N.~et al. 2003,  A\&A, 397, 147

\bibitem[{Ramsay \& Cropper(2004)}]{ramsay04a}
Ramsay, G., \& Cropper, M. 2004, MNRAS, 347, 497

\bibitem[{Ramsay et al.(2004)}]{ramsay04b}
Ramsay, G.~et al. 2004, MNRAS, 350, 1373 

\bibitem[{Ramsay et al.(2009)}]{ramsay09}
Ramsay, G.~et al. 2009, MNRAS, 395, 416

\bibitem[{Revnivtsev et al.(2004)}]{rev04}
Revnivtsev, M.G.~et al. 2004, Astronomy Letters, 30, 772

\bibitem[{Revnivtsev et al.(2009)}]{rev09}
Revnivtsev, M.G.~et al. 2009, Nature, 458, 1142 

\bibitem[{Ritter \& Kolb (2003}]{ritter03}
Ritter, H., \& Kolb, U. 2003, A\&A, 404, 301 (update RKcat7.16, 2011)

\bibitem[{Saxton et al.(2005)}]{saxton05}
Saxton, C.J.~et al. 2005, MNRAS, 360, 1091

\bibitem[{Saxton et al.(2007)}]{saxton07}
Saxton, C.J.~et al. 2007, MNRAS, 379, 779

\bibitem[{Scaringi et al.(2010)}]{scaringi10}
Scaringi, S.~et al. 2010, \mnras, 401, 2207

\bibitem[{Schwarz et al.(2005)}]{schwarz05}
Schwarz, R.~et al. 2005,  A\&A, 442, 271

\bibitem[{Schwope et al.(2007)}]{schwope07}
Schwope, A.D.~et al. 2007, A\&A, 469, 1027

\bibitem[{Schwope et al.(2009)}]{schwope09}
Schwope, A.D.~et al. 2009,  A\&A, 500, 867

\bibitem[{Still \& Mukai(2001)}]{still01}
Still, M. \& Mukai, K. 2001, \apj, 562, L71

\bibitem[{Suleimanov et al.(2005)}]{suleimanov05}
Suleimanov, V., Revnivtsev, M., Ritter, H. 2005,  A\&A, 435, 191

\bibitem[{Suleimanov et al.(2008)}]{suleimanov08}
Suleimanov, V..~et al. 2008, A\&A, 491, 525

\bibitem[{Szkody et al.(2004)}]{szkody04}
Szkody, P.~et al. 2004, Astron. J., 128, 2443  

\bibitem[{Traulsen et al.(2010)}]{traulsen10}
Traulsen, I.~et al. 2010, A\&A, 516, A76   

\bibitem[{Traulsen et al.(2011)}]{traulsen11}
Traulsen, I.~et al. 2011, A\&A, 529, 116   


\bibitem[{Vogel et al.(2007)}]{vogel07}
Vogel, J.,  Schwope, A.D., \& G\"ansicke, B.T. 2007,  A\&A, 464, 647

\bibitem[{Vogel et al.(2011)}]{vogel11}
Vogel, J., Schwope, A.D., \& Schwarz, R. 2011, A\&A, 530, A117

\bibitem[{Wickramasinghe \& Ferrario (2000)}]{wick00}
Wickramasinghe, D.T. \& Ferrario L. 2000, PASP, 112, 873

\bibitem[{Woelk \& Beuermann(1996)}]{woelk96}
Woelk, U., \& Beuermann, K. 1996, A\&A, 306, 232 

\bibitem[{Wu (2000)}]{wu00}
Wu, K. 2000, Space Science Rev., 93, 611

\bibitem[{Wu et al.(2001)}]{wu01}
Wu, K., et al. 2001, \mnras, 327, 208

\bibitem[{Yuasa et al.(2010)}]{yuasa10}
Yuasa, T.~et al. 2010, A\&A, 520, A25


\end{thebibliography}

\bigskip
\bigskip
\noindent {\bf DISCUSSION}

\bigskip
\noindent{\bf DAVID BUCKLEY} : 
Is it still the case that there is a lack of deeply eclipsing confirmed IPs, compared to polars and if so, why do you think this is the case?

\bigskip
\noindent {\bf MARTINE MOUCHET} : 
To my knowledge, there are no newly discovered eclipsing IPs. This might be 
related to the statistics or to their weak (diffused) X-ray emission if it
is partly eclipsed by the accretion disk. 
\end{document}